\documentclass[final,5p,times,twocolumn,sort&compress]{elsarticle}

\usepackage{multirow,setspace,amssymb,amsmath,graphicx,color,rotating,subfigure,url}
\usepackage{lineno}
\usepackage{natbib}
\usepackage{endfloat}
\usepackage{textcomp}

\journal{Physics Letters A} 

\begin{document}

\begin{frontmatter}

\title{Epidemic Spreading on Weighted Complex Networks}

\author{Ye Sun}
\author{Chuang Liu\corref{cor}}
\ead{liuchuang@hznu.edu.cn} %
\author{Chu-Xu Zhang}
\author{Zi-Ke Zhang\corref{cor}}
\cortext[cor]{Corresponding author.}
\ead{zhangzike@gmail.com} %

\address[AA]{Institute of Information Economy, Hangzhou Normal University, Hangzhou 311121, P. R. China}
\address[BB]{Alibaba Research Center of Complexity Science, Hangzhou Normal University, Hangzhou 311121, P. R. China}

\begin{abstract}
Nowadays, the emergence of online services provides various multi-relation information to support the comprehensive understanding of the epidemic spreading process. In this Letter, we consider the edge weights to represent such multi-role relations. In addition, we perform detailed analysis of two representative metrics, outbreak threshold and epidemic prevalence, on SIS and SIR models. Both theoretical and simulation results find good agreements with each other. Furthermore, experiments show that, on fully mixed networks, the weight distribution on edges would not affect the epidemic results once the average weight of whole network is fixed. This work may shed some light on the in-depth understanding of epidemic spreading on multi-relation and weighted networks.
\end{abstract}

\begin{keyword}
Complex networks; Epidemic spreading; Multi-relation; Weighted network
\end{keyword}

\end{frontmatter}

\section{\label{S1:Intro}Introduction}

Epidemic spreading based on complex networks, where nodes represent individuals and links denote their interactions, has attracted an increasing attention in recent years \cite{Van-Mieghem-Van-de-Bovenkamp-2013-PRL,Castellano-Fortunato-Loreto-2009-RMP,Newman-2002-PRE}. Generally, disease propagation can be modeled as a kind of dynamic process in which an item is transmitted from an infected individual to a susceptible individual via the link between them \cite{Lloyd-May-2001-Science}. Motivated by previous pioneering works that many real networks exhibit the small-world phenomenon and scale-free property, more and more results of spreading dynamics on those networks are presented recently \cite{Zhou-Fu-Wang-2006-PNS,Liu-Zhang-CNSNS-2013}. The spreading process on the scale-free network indicates that a highly heterogeneous structure would lead to both the absence of the epidemic threshold \cite{Pastor-Satorras-Vespignani-2001-PRL,Pastor-Satorras-Vespignani-2001-PRE} and the hierarchical spreading of epidemic outbreak \cite{Barth-Barrat-Pastor-Satorras-Vespignani-2004-PRL}. Further study of the susceptible-infected-susceptible (SIS) model on the scale-free network shows that the vanishing of epidemic threshold stems from the node with the largest degree rather than the scale-free nature \cite{Castellano-Pastor-Satorras-2010-PRL}. More general, the epidemic threshold for SIS model on an arbitrary undirected graph is determined by the largest eigenvalue of the adjacency matrix \cite{Gomez-Arenas-BorgeHolthoefer-Meloni-Moreno-2010-EPL,Mieghem-2013-EPL}. On the small-world network, most infection occurs locally for the high-level cluster and the disease spreads rapidly into large regions of the population for the short path lengths \cite{Keeling-Eames-2005-JRSI,Kuperman-Abramson-2001-PRL}. Analysis of the susceptible-infected-recovered model (SIR) on small-world networks presents that a phase transition between two different regimes occurs at a particular rewiring parameter $p_c$ \cite{Zanette-2001-PRE}, and such critical transition is also found in the spreading on dynamical small-world networks \cite{Stone-Mckay-2011-EPL}. In addition, the epidemic propagation on the real-network structure also draws much attention, such as sexually transmitted disease on the sexual contact networks \cite{Gomez-Gardenes-Latora-Moreno-Profumo-2008-PNAS}, mobile phone viruses on the multimedia messaging systems \cite{Wang-Gonzalez-Hidalgo-Barabasi-2009-Science}, disease transmission  between human beings and mosquitos \cite{WangY2012}, and so on.

However, the aforementioned researches mostly consider the simplest case of networks with only one type of links. In fact, there exist various real-world complex networks, which are characterized by inherent multi-relation connections \cite{Szell-Lambiotte-Thurner-2010-PNAS,Buldyrev-Parshani-Paul-Stanley-Havlin-2010-Nature,Li-Zou-Guan-Gong-Li-Di-Lai-2013-Arxiv}, such as blood relationship, romantic relationship, friend relationship, work relationship in the social contact network. The role of hybrid relations in the spreading process could be very different \cite{Yagan-Gligor-2012-PRE}. Some disease propagation would be more likely to be promoted among family members such as the HIV, while some contagions such as H7N9 \cite{Uyeki-Cox-2013-NEJM} are prone to transmit among the staffs in the slaughter house or chicken farm. It is obvious that with the existence of the multiple relationship, the network structure becomes more complex and diverse, leading to more special spreading dynamics. Failure cascading of the network coupled with connectivity links and dependency links \cite{Parshani-Buldyrev-Havlin-2011-PNAS,Son-Bizhani-Christensen-Grassberger-Paczuski-2012-EPL,Brummitt-Lee-Goh-2012-PRE} demonstrates that the network disintegrates in a form of a first-order phase transition for a high density of dependency links, whereas the network disintegrates in a second-order transition for a low density of dependency links. Though multi-relation networks attract more and more attention, it is still unclear
how the multi-relationship affects the  epidemic spreading dynamic for the complex network structure. It is a reasonable way to treat the multi-relation network with assigning different weights for each relation. Li \emph{et al}~\cite{Li-Xu-Tang-2013-APS} proposed a binary-relation network model, representing colleague and friend relationship by setting different weights of the corresponding links, and the epidemic spreading process demonstrates that the outbreak threshold is suppressed by the closer relationship.

\begin{figure}
\centering
\includegraphics[width=8cm]{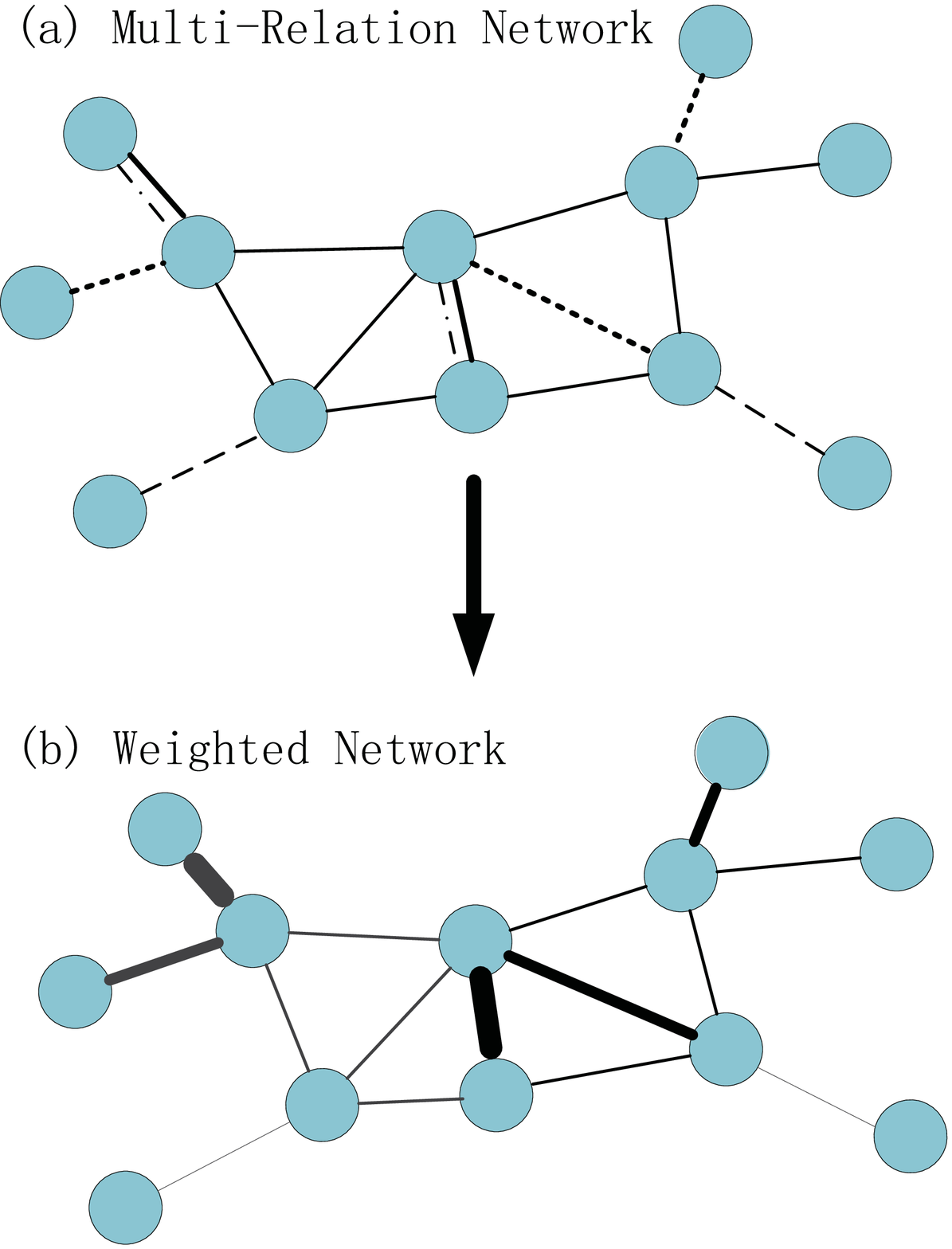}
\caption{(Color online)\label{Fig:Illus} Illustration of (a) multi-relation network where each type of line corresponds to one kind of relation; and (b) the corresponding weighted network, where the thickness of link represents the size of weight. }
\end{figure}

In order to understand the epidemic spreading process on multi-relation networks in-depth, in this Letter, we construct multi-relation networks with considering various relation-levels as different weights, where link weight follows some given distributions (see Fig. \ref{Fig:Illus}). Then, we perform SIS and SIR models on the proposed weighted networks, where the links with the same weight shows the same transmit capacity. Focusing on the outbreak threshold and epidemic prevalence, theoretical analysis based on the mean-field approximation illustrates that multiple relations would result in the decrease of the outbreak threshold and brings more infections in the final state. Detailed analysis indicates that the epidemic spreading result just depends on the average level of relationship rather than the link weight distribution. In addition, Monte Carlo simulation agrees well with the theoretical results.

\section{Model}

In this Letter, we consider that there are $n$ kinds of relations in the network where the multiple relations can be represented as the link weight. Fig. \ref{Fig:Illus} shows such a illustration of a multi-relation network and its corresponding weighted network. In general, we set the links with discrete weights as $w=1,2,...,n$ to identify each type of relationship, and the link with higher weight means closer relationship, through which disease is more likely to transmit. In order to illustrate the spreading effects of different relations, we investigate two sets of weight distributions, one of which follows uniform distribution, the other follows the poisson distribution. In addition, we assume that all links are fully mixed and the same type of links are distributed uniformly in the network.

Consequently, we adopt SIS and SIR models on the small-world network (WS network with randomness probability $p_s=0.3$) \cite{Watts-Strogatz-1998-Nature} and scale-free network (BA network) \cite{Barabasi-Albert-1999-Science}, respectively, where the network size is set to be $N=10^4$, and the average degree is $\langle k\rangle=8$. In general, we set the recovery probability $\mu=1$, initial infected density $I_0=0.01$ and define the transmit probability for links with $w=1$ as $\lambda$. We assume that transmit probability through the edge with weight $w$ ($\lambda_w$) is equivalent to the infected probability that $w$ infected individuals (I) simultaneously influence the susceptible individual (S) \cite{Tasgin-Bingol-2012-ACS}, which can be obtained by :
\begin{equation}
\lambda_{w}=1-(1-\lambda)^{w}.
\end{equation}

According to the mean-field approximation, for an arbitrary edge the successful transmission probability in one timestep is:
\begin{equation}
\beta=\sum_w p_w(1-(1-\lambda)^w),  \label{eq:beta1}
\end{equation}
where $p_w$ is the proportion of links with weight $w$.

In general, $\lambda$ is very small, thus Eq. (\ref{eq:beta1}) can be simplified to:
\begin{equation}
\beta \approx \alpha\lambda,  \label{eq:beta2}
\end{equation}
where $\alpha$ is the average weight of all links in the network.


\begin{figure}
\centering
\includegraphics[width=8cm]{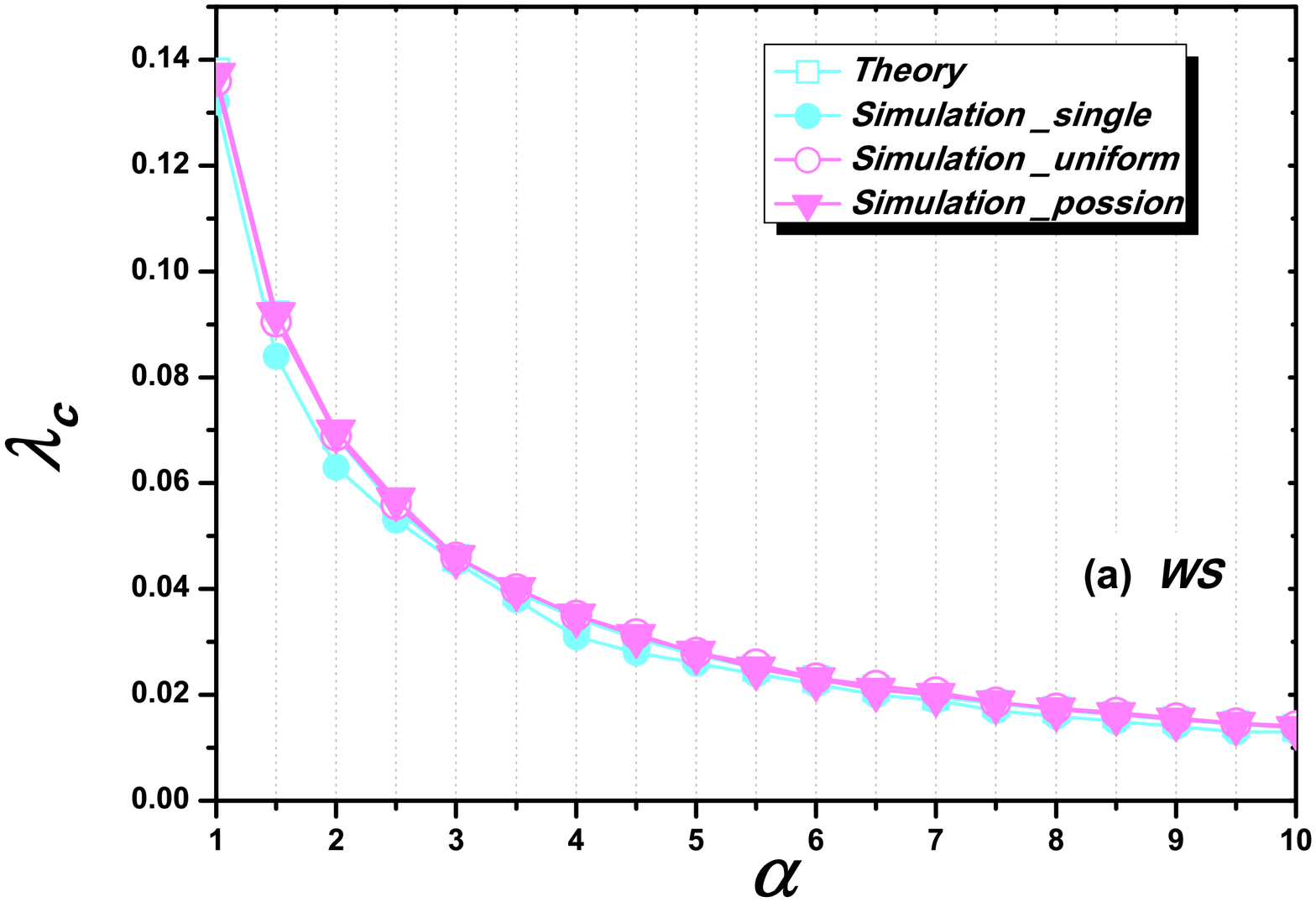}
\includegraphics[width=8cm]{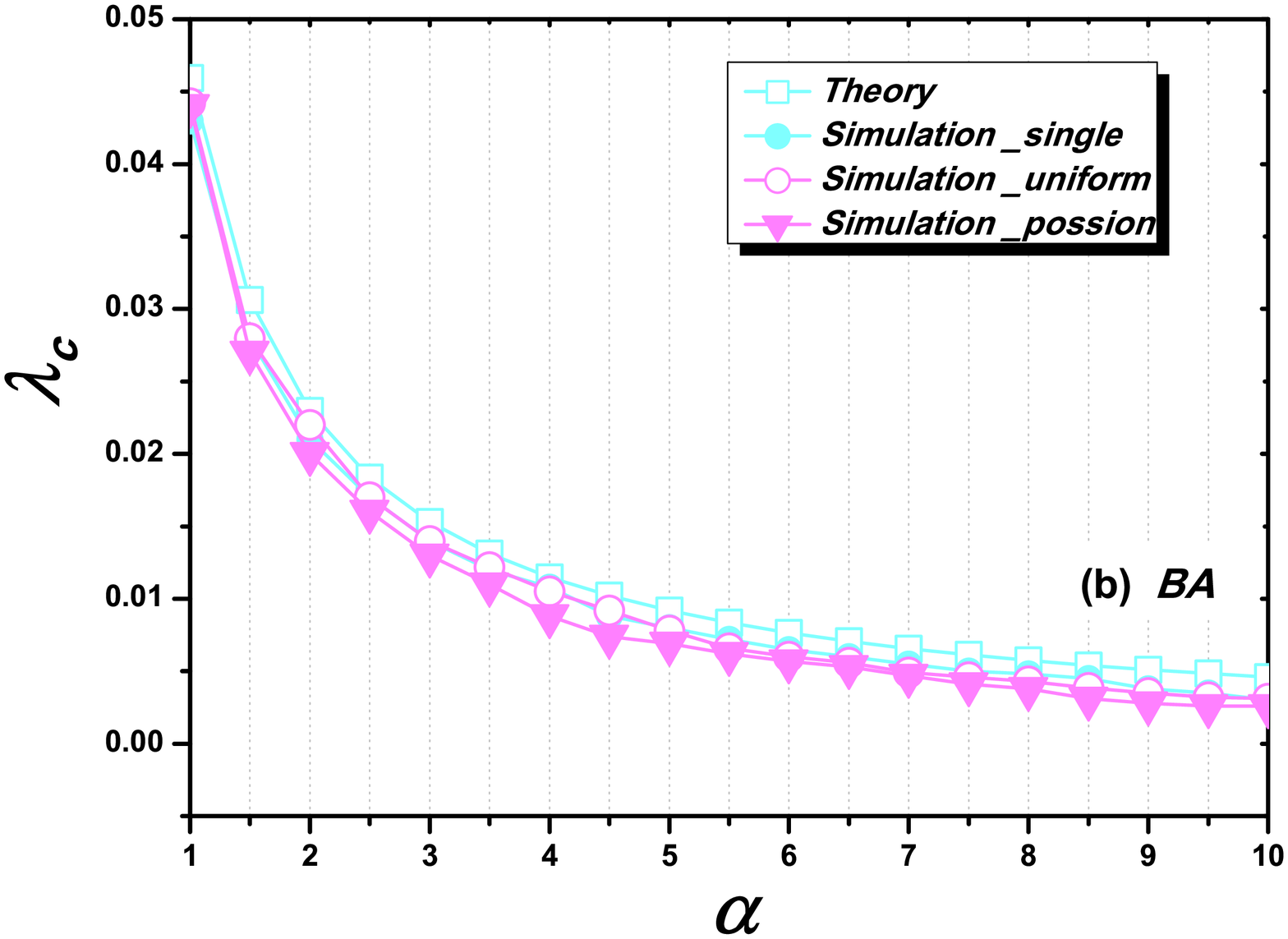}
\caption{(Color online)\label{Fig:Threshold:SIR} Epidemic threshold $\lambda_c$ as a function of the parameter $\alpha$ on the SIR model for (a) WS network; (b) BA network. }
\end{figure}

\section{Outbreak Threshold}

\begin{figure}
\centering
\includegraphics[width=8cm]{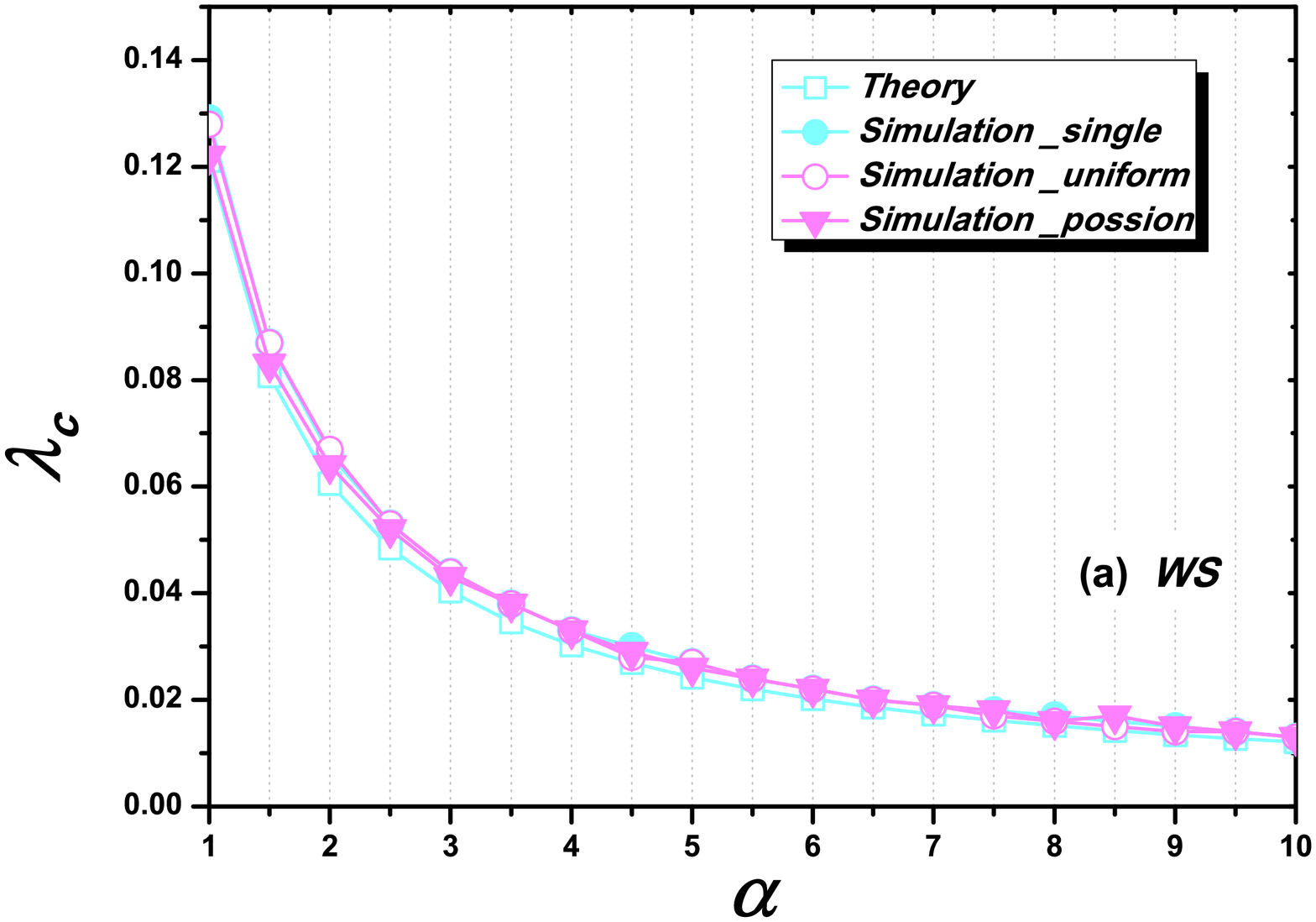}
\includegraphics[width=8cm]{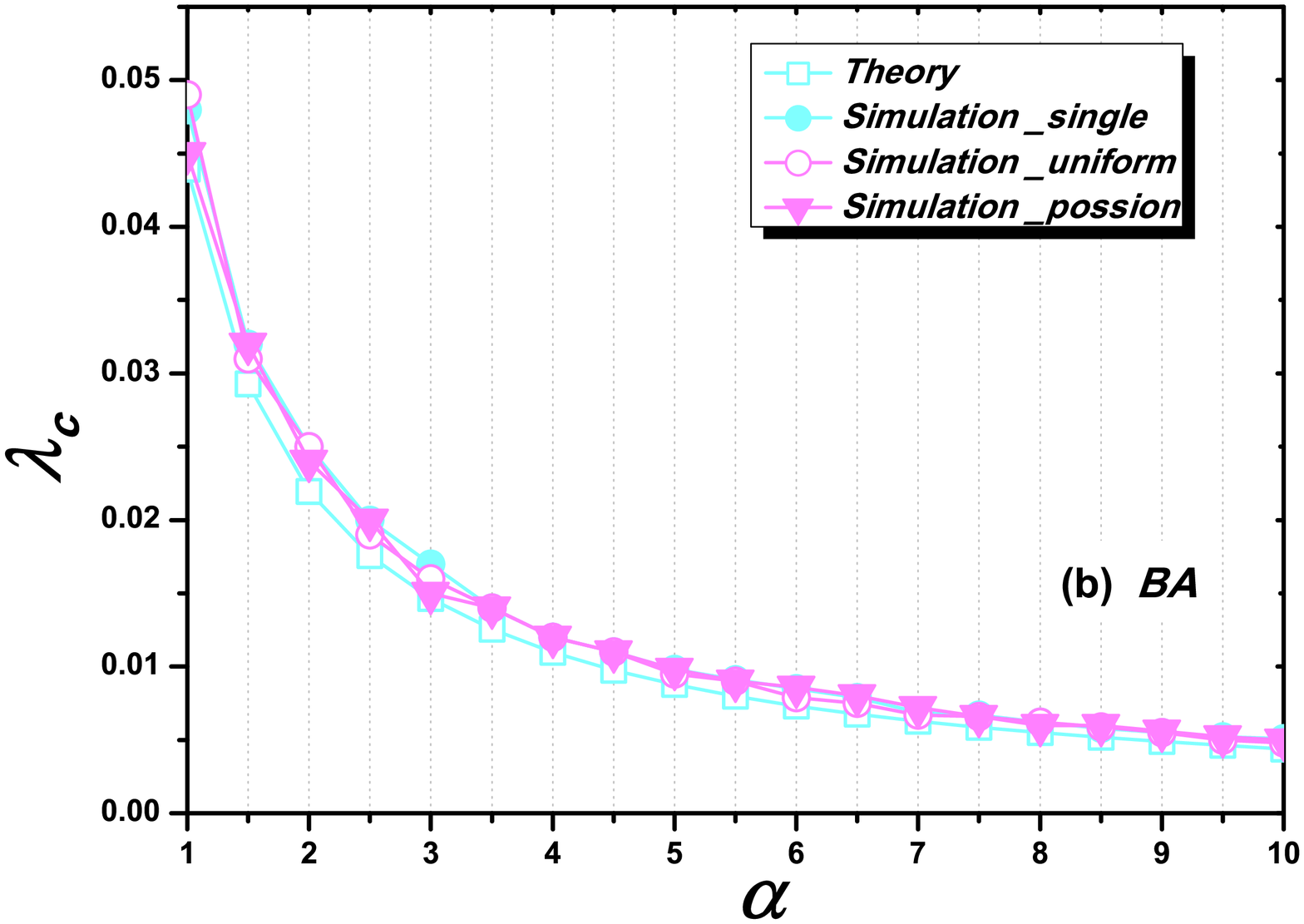}
\caption{(Color online)\label{Fig:Threshold:SIS} Epidemic threshold $\lambda_c$ as a function of the parameter $\alpha$ on the SIS model for (a) WS network; (b) BA network. }
\end{figure}

In order to understand the epidemic outbreak threshold with the multi-relation effect, we use a method of percolation theory, for disease spreading can be seen as a growing percolation process \cite{Parshani-Carmi-Havlin-2010-PRL}. For the case of uncorrelated networks, the probability that an edge links to a node with degree $k$ is $\frac{\displaystyle kp(k)}{\displaystyle \langle k\rangle}$, where $p(k)$ is the degree distribution of the observed network, and $\langle k\rangle$ is the average degree. In addition, we assume that as long as the epidemic has not spread out yet, the infected node with degree $k$ has only one ingoing link and $k-1$ outgoing links \cite{Parshani-Carmi-Havlin-2010-PRL}. And the average number of susceptible nodes infected by an already infected node $i$ is:
\begin{align}
\langle n_{i}\rangle=\beta\sum_{k}\frac{p(k)k(k-1)}{\langle k\rangle}+\pi,\label{eq:ni}
\end{align}
where $\pi$ is the contribution of the probability to reinfect the ancestor (the node that infected node $i$, corresponding to $i$'s ingoing link) \cite{Madar-Kalisky-Cohen-ben-Avraham-Havlin-2004-EPJB} .

For the \textbf{SIR} model where the reinfection is forbidden, the disease spreads directionally down a tree structure and $\pi=0$. If an infected individual infects at least one other individual on average, the epidemic can reach an endemic state. Therefore, we have $\langle n_i\rangle=1$ at the epidemic threshold \cite{Parshani-Carmi-Havlin-2010-PRL,Madar-Kalisky-Cohen-ben-Avraham-Havlin-2004-EPJB}, leading to:
\begin{equation}
\lambda_{c}(SIR)=\frac{\langle k\rangle}{\alpha(\langle k^{2}\rangle-\langle k\rangle)}. \label{eq:sir:threshold}
\end{equation}

For the \textbf{SIS} model where the reinfection is allowed, things get more complicated. We define $\pi_t$ as the probability that $j$ infects $i$ if $i$ has infected $j$ yet. In this model, we set the recovery probability $\mu=1$, which means that the infected node remains the infected state in just one step. Therefore, the interval that $j$ remains infected and $i$ remains susceptible is only 1, which leads to $\pi_t=\beta$. Incorporating the effect of competition between $j$ and the other descendants of $i$, the reinfected probability $\pi$ for the system can be calculated as following \cite{Parshani-Carmi-Havlin-2010-PRL}:
\begin{equation}
\pi=\pi_{t}\sum^{\kappa-1}_{k'=0}\left( {\begin{array}{*{20}c}
   \kappa-1  \\
   k'  \\
\end{array}} \right)\frac{(\beta\pi_{t})^{k'}(1-\beta\pi_{t})^{\kappa-1-k'}}{k'+1}, \label{eq:cggl}
\end{equation}
where $\kappa-1=\sum\limits_{k}\frac{\displaystyle p(k)k(k-1)}{\displaystyle \langle k\rangle}$ is the branching factor that represents the average number of nodes influenced by node $i$ and $k'$ represents the infected neighbors of node $i$.

Neglect the high-order term, we can obtain that $\pi \approx \beta$. According to Eq. (\ref{eq:ni}), the epidemic threshold of SIS model is:
\begin{equation}
\lambda_{c}(SIS)=\frac{\langle k\rangle}{\alpha\langle k^{2}\rangle}. \label{eq:sis:threshold}
\end{equation}

\begin{figure}
\centering
\includegraphics[width=8cm]{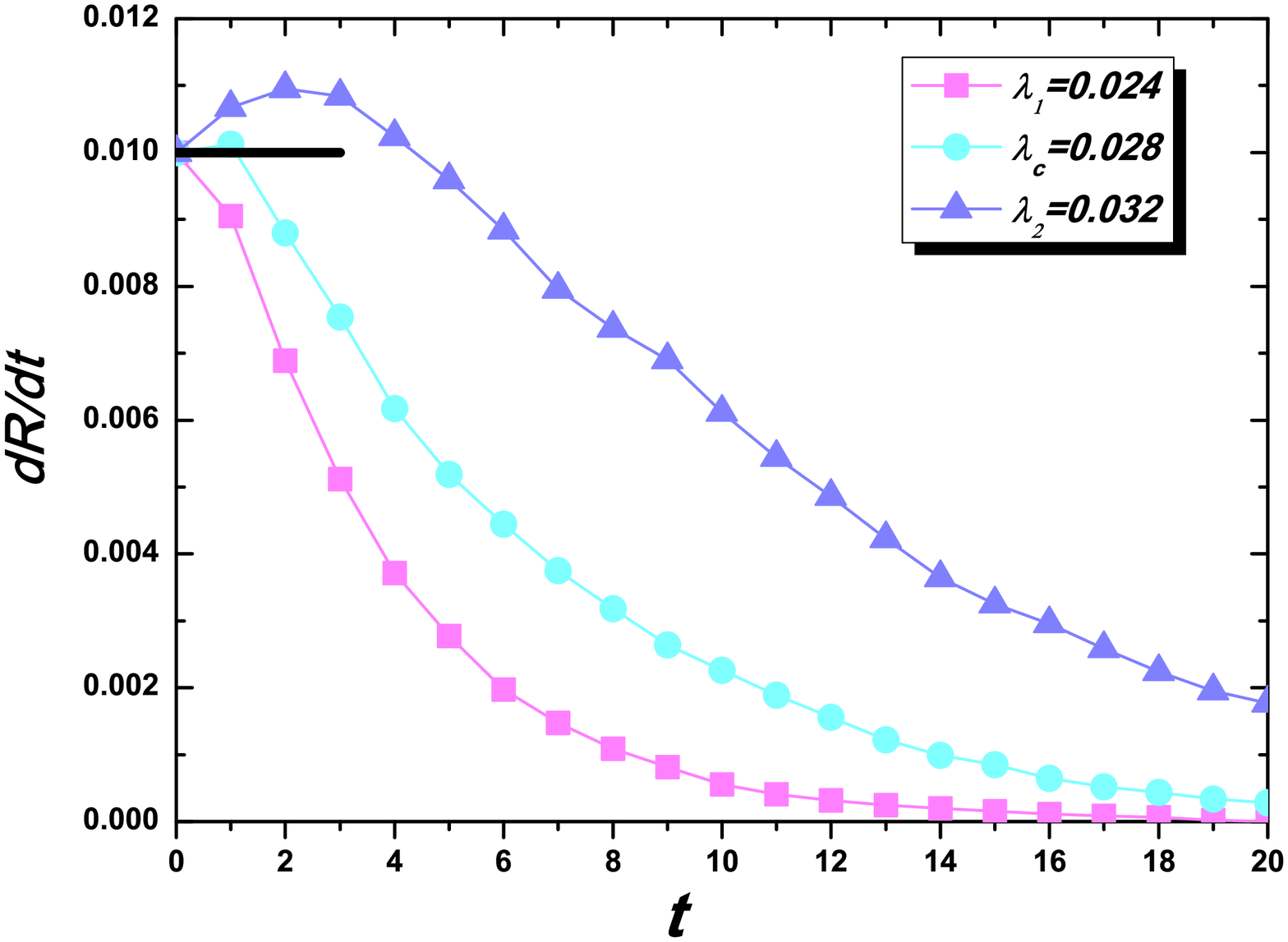}
\caption{(Color online)\label{Fig:Explain:SIR} Illustration of the method that locates the threshold value in the SIR model. }
\end{figure}

Fig.~\ref{Fig:Threshold:SIR} and Fig.~\ref{Fig:Threshold:SIS} illustrate the theoretical analysis and simulation results of the epidemic outbreak threshold on multi-relation networks for SIR and SIS models, respectively. The theoretical results are obtained according to Eq. (\ref{eq:sir:threshold}) and Eq. (\ref{eq:sis:threshold}), where the parameters such as $\langle k\rangle$, $\langle k^{2}\rangle$ can be obtained from the corresponding synthetic networks. For the Monte Carlo simulations, we synthesize the underlying network with setting the multi-relation links. At the beginning, we randomly select $NI_0$ nodes as the epidemic \textit{seeds}. However, for the simulation of SIR model, due to the random fluctuations, it is not easy to precisely determine the epidemic outbreak threshold. In this Letter, we define $dR/dt$ as the new \emph{R}-state nodes in each timestep, and the outbreak threshold is the $\lambda$ value that the trend of $dR/dt$ changes from ``monotonic decrease" to ``rise first and then fall", just as the circle symbols in Fig. \ref{Fig:Explain:SIR}. In Fig. \ref{Fig:Explain:SIR}, we plot $dR/dt$ versus time step $t$ in WS network as an example to show the threshold value in the SIR model, where $\alpha = 5$ and $<k> = 8$. According to Eq. (\ref{eq:sir:threshold}), we can obtain that the outbreak threshold is round 0.028. When $\lambda=0.024$ (square symbols), $dR/dt$ decreases monotonously for the spread can¡¯t spread out. when $\lambda=0.032$ (triangle symbols), $dR/dt$ rises for a few steps and then decreases. And the critical point is just corresponding to the trend of $dR/dt$ changes between the two states. For the simulation of SIS model, the outbreak threshold is the $\lambda$ value that the number of infected nodes in the final state changes from zero to nonzero. Simulation results are obtained by averaging over 100 independent realizations.

Based on the mean-field approximation, theoretical analysis illustrates that the outbreak threshold is influenced by the average edge weight rather than the weight distribution. Simulation result is consistent with theoretical analysis very well. In order to confirm this, besides the uniform and poisson distribution, we also give the simulation of the network that just with single relationship, that is to say, all edge weights are identical to $\alpha$. Simulation results (see Fig. \ref{Fig:Threshold:SIR} and Fig. \ref{Fig:Threshold:SIS}) present that, when those three strategies share the same average weight $\alpha$, very similar outbreak thresholds will be obtained. It is noted that the weight heterogeneity is very different for the three considering distributions, but the simulation results show that the weight distribution effect is very weak, which is at odds with epidemic research on the weighted network where the epidemic spreading only transmits to one neighbor with the contact probability determined by the edge weight \cite{Yang-Zhou-2012-PRE} .

The dependence of the threshold on parameter $\alpha$ presents that epidemic threshold decreases with the increase of $\alpha$. Large $\alpha$ displays that there are more close relationships in the system, which indicates that the closer relationships suppresses the threshold. Comparing different underlying networks, the influence of the multi-relations is more obvious on WS network than that on BA network. This may be caused by that the hub nodes in the heterogeneous network will sustain activity even with very small transmit probability \cite{Castellano-PastorSatorras-2012-SR}. For a long-tailed degree distribution with power-law form in BA network, $\frac{\displaystyle \langle k\rangle}{\displaystyle \langle k^2\rangle} \ll 1$ which represents the diminishing outbreak threshold in the thermodynamic limit \cite{Castellano-Pastor-Satorras-2010-PRL} according to Eq. (\ref{eq:sir:threshold}) and Eq. (\ref{eq:sis:threshold}) even with very small $\alpha$. Therefore, introducing the close relationships leads to limited enhancement in BA network.

\section{Epidemic Prevalence}

\begin{figure}
\centering
\includegraphics[width=8cm]{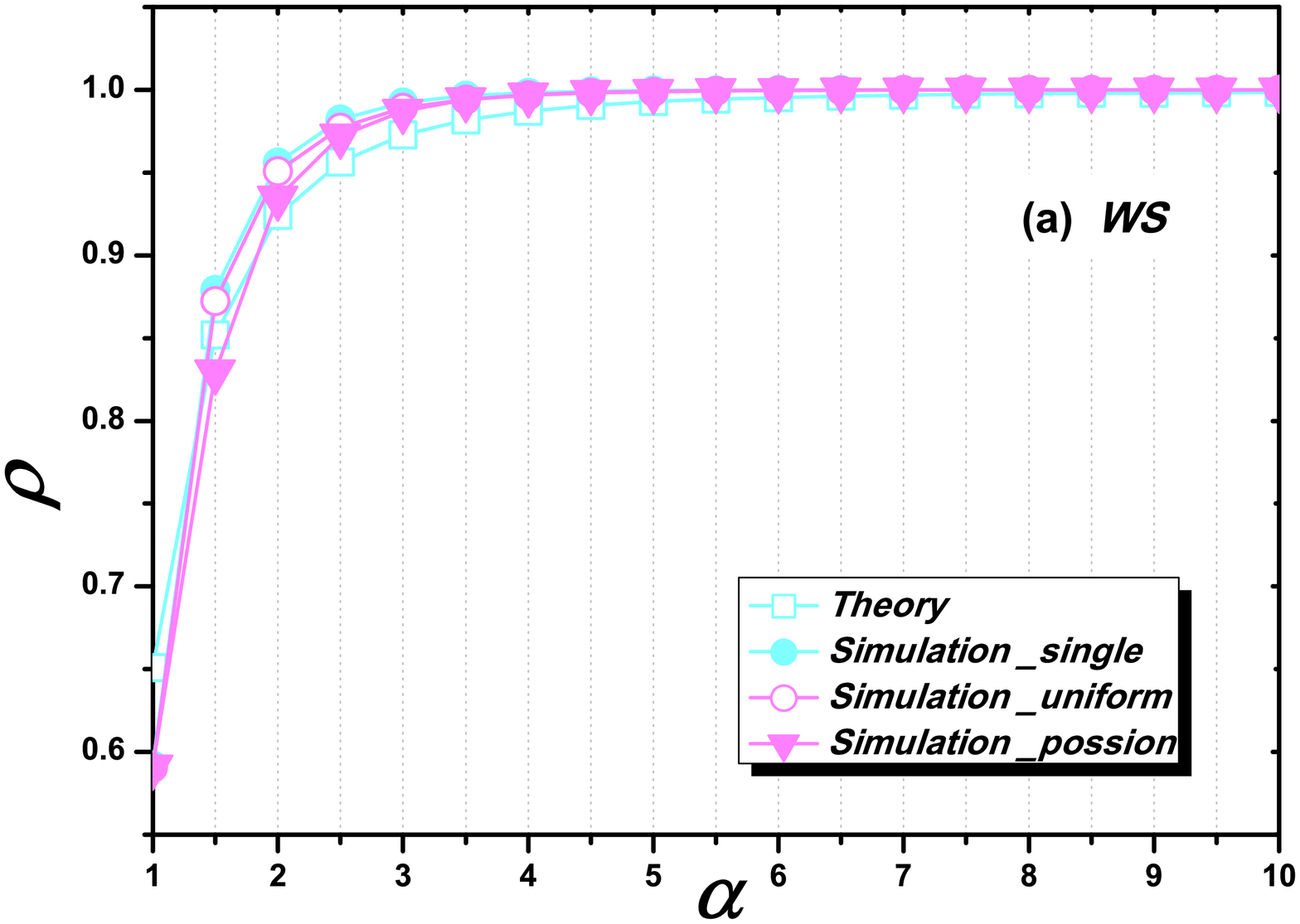}
\includegraphics[width=8cm]{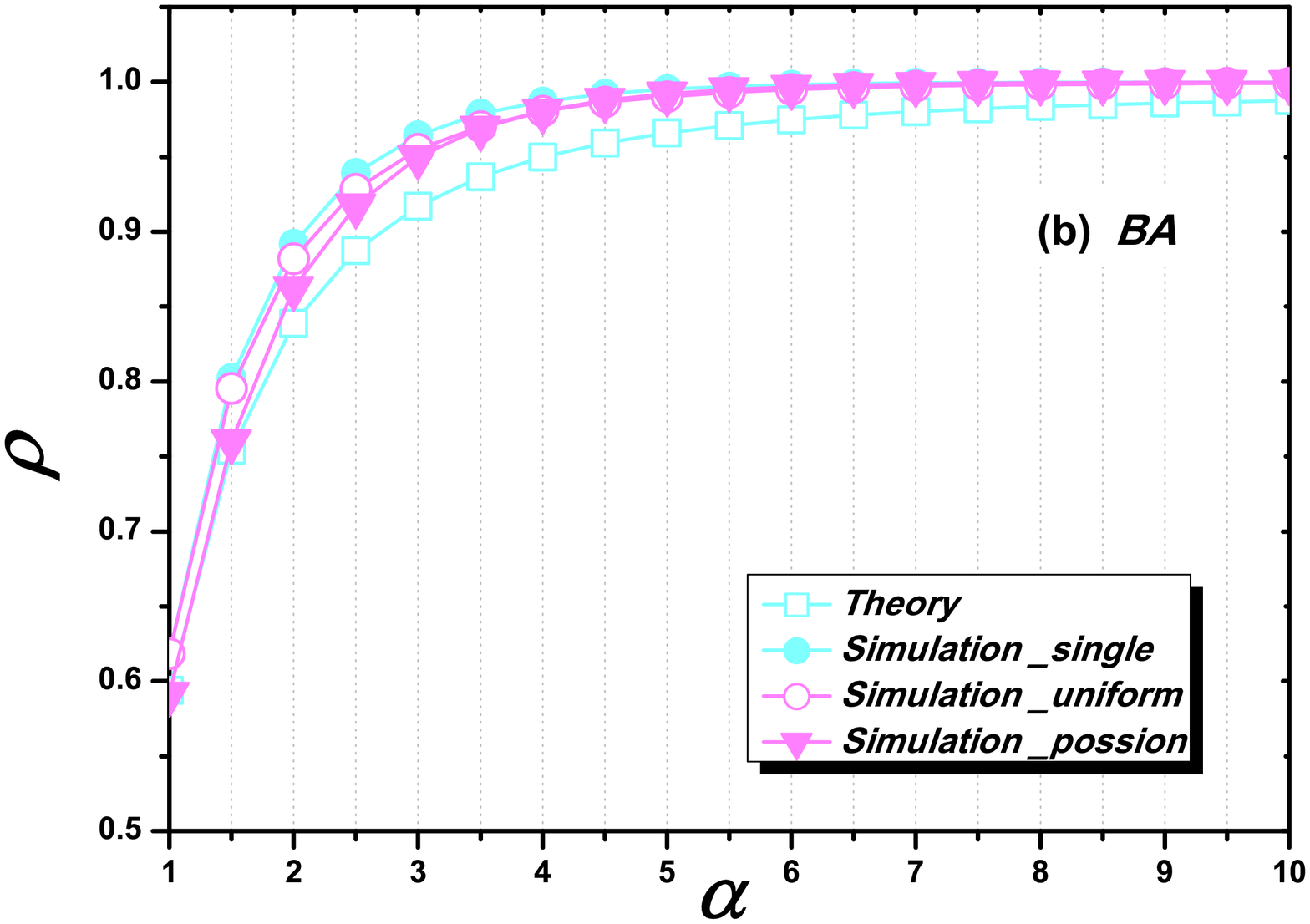}
\caption{(Color online)\label{Fig:Density:SIR} Epidemic prevalence $\rho$ as a function of the parameter $\alpha$ on the SIR model for (a) WS network; (b) BA network.}
\end{figure}

In the \textbf{SIR} model, we use the density of the recovered nodes in the final state to illustrate the epidemic prevalence. We denote $S_k(t)$, $I_k(t)$ and $R_k(t)$ as the fraction of the susceptible, infected and recovered individuals with degree $k$ at time $t$. At the mean-field level, these densities satisfy the following differential equations \cite{Moreno-Pastor-Satorras-Vespignani-2002-EPJB}:
\begin{equation}
\renewcommand{\arraystretch}{1.8}
\left\{
\begin{array}{*{30}l}
   \frac{\displaystyle d S_{k}(t)}{\displaystyle dt}=-\beta kS_{k}(t)\Theta(t)  \\
   \frac{\displaystyle d I_{k}(t)}{\displaystyle dt}=\beta kS_{k}(t)\Theta(t)-I_{k}(t),  \\
   \frac{\displaystyle d R_{k}(t)}{\displaystyle dt}=I_{k}(t)
\end{array}
\right.
\end{equation}
where $\Theta(t)=\sum\limits_k \frac{ \displaystyle kp(k)I_k(t)}{\displaystyle \langle k\rangle}$, represents the probability that an arbitrary link points to an infected node.

With the direct integration of $\frac{\displaystyle d S_{k}(t)}{\displaystyle dt}$, we can obtain that:
$S_{k}(t)=S_{0}e ^{-\beta k\phi(t)}$, where $\phi(t)=\sum\limits_{k}\frac{\displaystyle kp(k)R_{k}(t)}{\displaystyle \langle k\rangle}$, and $S_0=1-I_0=0.99$ is the initial susceptible density.

Through the differential equation for $\phi(t)$:
\begin{equation}
\frac{d\phi(t)}{dt}=1-\phi(t)-\sum\limits_{k}\frac{kp(k)S_{0}e ^{-\beta k\phi(t)}}{\langle k\rangle}. \label{eq:mean}
\end{equation}

For the condition that $\frac{\displaystyle d\phi(t)}{\displaystyle dt}=0$ when $t\rightarrow \infty$, we can obtain $\phi_{\infty}$ ,
\begin{equation}
\phi_{\infty}=1-\sum\limits_{k}\frac{kp(k)S_{0}e^{-\beta k\phi_{\infty}}}{\langle k\rangle}.\label{eq:phi}
\end{equation}

Once we obtain the numerical solution of $\phi_{\infty}$, we can calculate the total epidemic prevalence:
\begin{equation}
\rho_R=R_{\infty}=1-S_{\infty}=1-\sum\limits_{k}p(k)S_{0}e ^{-\beta k\phi(\infty)}. \label{eq:sir:prevalence}
\end{equation}

In the \textbf{SIS} model, we use the density of the infected nodes in the final state to illustrate the epidemic prevalence. In heterogeneous mean-field theory, it is supposed that all nodes of the same degree share the similar dynamic behavior. Based on the microscopic Markov-chain approach \cite{Gomez-GomezGardenes-Moreno-Arenas-2011-PRE}, we can obtain the infected density of node degree $k$, $I_k$, for $t\rightarrow \infty$ as follows:
\begin{equation}
I_{k}=(1-I_{k})(1-\vartheta_k)+(1-\mu)I_k,\label{eq:chafen}
\end{equation}
where $\vartheta_k$ is the probability that the node with degree $k$ are not being infected by any neighbors, which can be obtained by:

\begin{equation}
\vartheta_{k}=\prod_{k'}^{M}(1-\beta I_{k'})^{C_{kk'}}, \label{eq:chafen2}
\end{equation}
where $C_{kk'}=kp(k'|k)=\frac{\displaystyle kk'p(k')}{\displaystyle \langle k\rangle}$, represents the expected number of links from a node of degree $k$ to nodes of degree $k'$.

Then the global epidemic prevalence can be obtained as follows:
\begin{equation}
\rho_I=\sum_{k}p(k)I_{k}. \label{eq:sis:prevalence}
\end{equation}

\begin{figure}
\centering
\includegraphics[width=8cm]{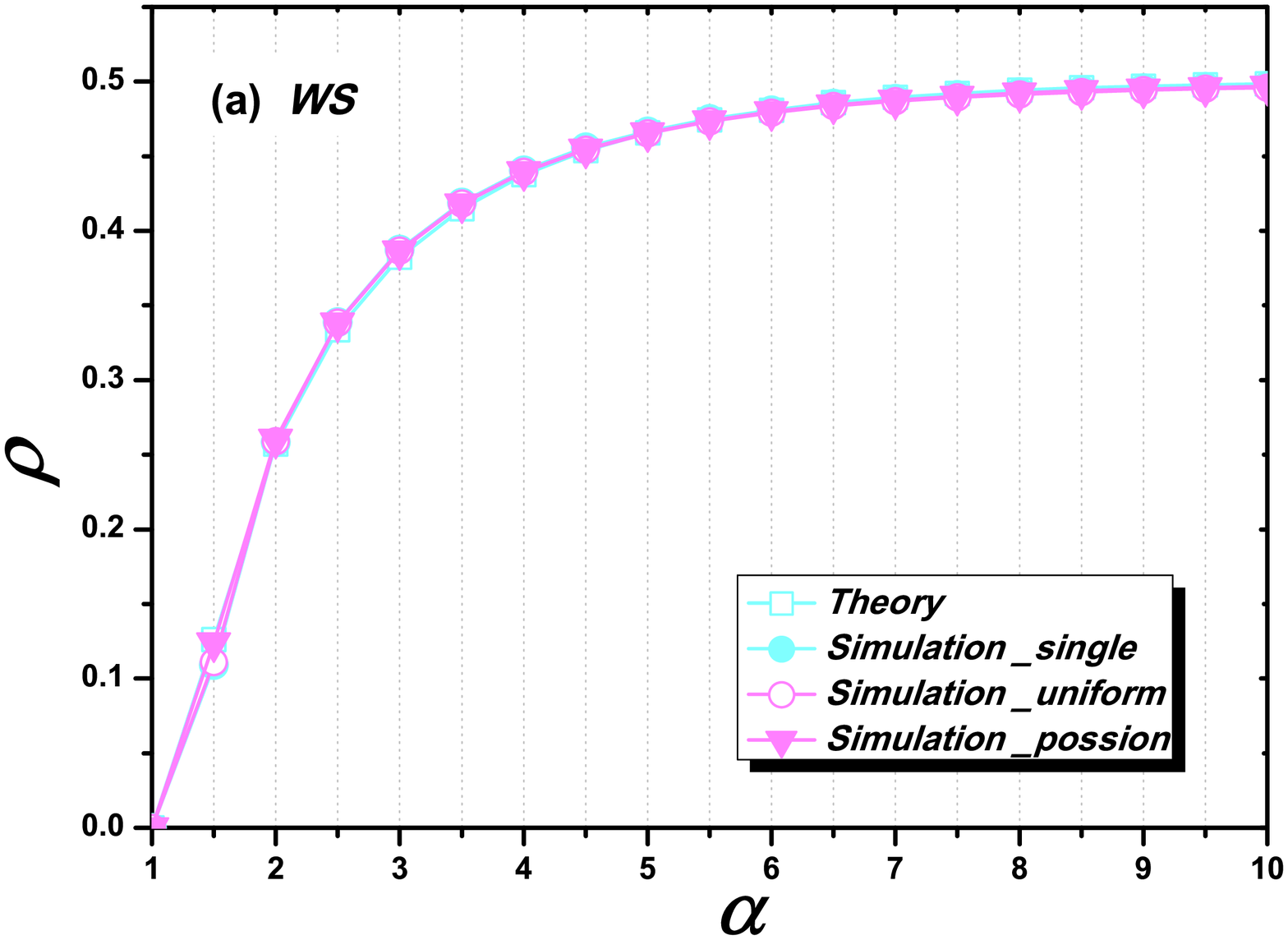}
\includegraphics[width=8cm]{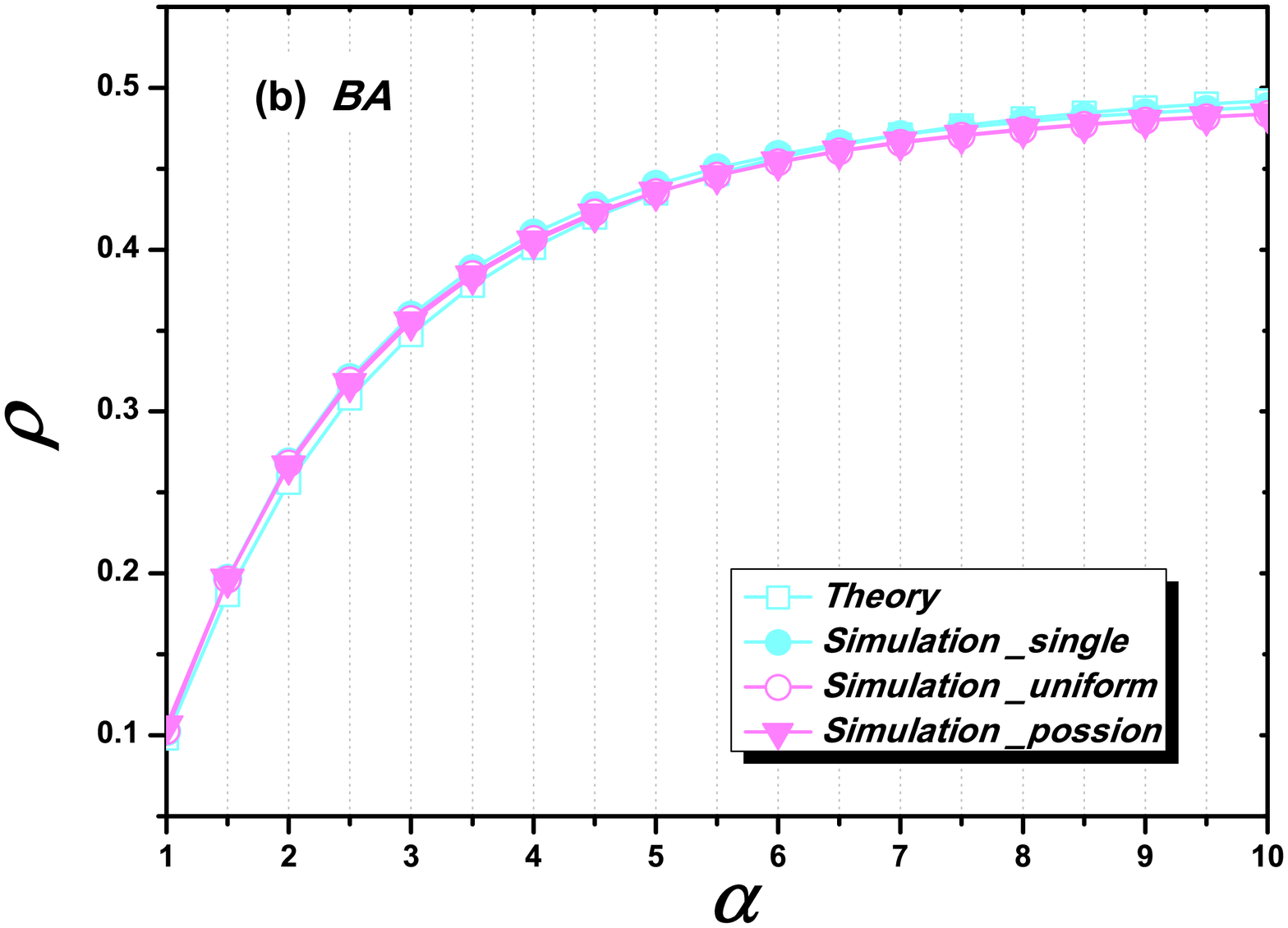}
\caption{(Color online)\label{Fig:Density:SIS} Epidemic prevalence $\rho$ as a function of the parameter $\alpha$ on the SIS model for (a) WS network; (b) BA network. }
\end{figure}

Fig. \ref{Fig:Density:SIR} and Fig. \ref{Fig:Density:SIS} illustrate the epidemic prevalence as a function of the parameter $\alpha$ for SIR and SIS models, respectively. Theoretical results are obtained according to Eq. (\ref{eq:sir:prevalence}) and Eq. (\ref{eq:sis:prevalence}). Other parameters such as $p(k)$, $\langle k\rangle$ are calculated from the corresponding synthetic networks. Note that, $\phi_\infty=0$ is always a solution for Eq. (\ref{eq:phi}). In order to have a non-zero solution, the condition that $\lambda>\lambda_c(SIR)$ must be fully satisfied. In simulations, we set $\lambda=0.2$ for SIR and $\lambda=0.1$ for SIS models, respectively. In addition, the epidemic prevalence is defined as the density of the recovered nodes and the infected nodes at the final state for SIR and SIS models, and all simulation results are obtained by averaging over 100 independent realizations.

The theoretical analysis based on the mean-field approach, shown by Eq. (\ref{eq:sir:prevalence}) and Eq. (\ref{eq:sis:prevalence}), suggests that the epidemic prevalence is dependent on the average edge weight rather than weight distribution. In addition, simulation results show that the values of epidemic prevalence are identical for uniform and possion distribution when they have the same average weight $\alpha$, which finds good agreement with the theoretical results for both SIR and SIS models (see Fig. \ref{Fig:Density:SIR} and Fig. \ref{Fig:Density:SIS}). The small deviation between the simulation and theoretical results of different weight distribution indicates that the theoretical analysis based on the mean-field approximation is quite reasonable.

It also can be seen from Fig. \ref{Fig:Density:SIR} and Fig. \ref{Fig:Density:SIS} that the edge weight indeed enhance the epidemic prevalence. According to Eq. (\ref{eq:beta2}), larger $\alpha$ corresponds higher transmission probability, hence more individuals would be infected. Comparing with BA network, such enhancement is more significant on WS network, which might be caused by the network homogeneity. In WS network, strong links (correspond to large weight) would be distributed uniformly for the homogeneous network structure, which leads to a more global spreading process \cite{Pajevic-Plenz-2012-NP}. Comparatively, for BA network, such strong links would \emph{sit} around the central nodes with large probability, which just enhances the effect of local spreading within the cluster initiated by hub nodes. For the extreme case of SIS model, when $\alpha=1$ (correspond to unweighed condition), there is no infected node ($\rho_I=0$) at the final state on WS network,  while $\rho_I\approx 0.1$ for BA network (Fig. \ref{Fig:Density:SIS}). This is due to the reason that the parameter $\lambda=0.1$ is smaller than the outbreak threshold $\lambda_c=0.13$ (see Fig. \ref{Fig:Threshold:SIS}a). Furthermore, a relatively small $\alpha$ would much enhance the epidemic prevalence while deceasing the outbreak threshold on WS network. Therefore, link operations \cite{Bishop-Shames-2011-EPL} for immunizing the closer relational links would be an efficient approach to restrain the epidemic spreading on WS network. 

\section{Conclusions and Discussion}
In this Letter, we use the edge weight to study the dynamics of epidemic spreading on multi-relation networks. In addition, we perform both theoretical and simulation on two representative epidemic spreading models: SIS and SIR models, and find good agreements on the analysis of both outbreak threshold and epidemic prevalence. In addition, experiments show that, on fully mixed networks, the weight distribution on edges would not affect the epidemic results once the average weight of whole network is fixed, which is at odds with our common knowledge and other studies on single-contact spreading precess.

The findings of this work may have a wide-range application in studying the epidemic spreading. Considering that edge weights might change due to technology in the social connection, the spreading process can be simulated with the proposed method for it just depends on the average weights. The present work just provides a start point to study of the dynamics of weighted epidemic spreading process, more comprehensive and in-depth understanding and application need further efforts to discover.

\bigskip
{\textbf{Acknowledgments:}}

This work was partially supported by the National Natural Science Foundation of China (Grant Nos. 11105024, 11105040, 1147015, 11301490 and 11305043), and the Zhejiang Provincial Natural Science Foundation of China (Grant Nos. LY12A05003 and LQ13F030015), the start-up foundation and the Pangdeng project of Hangzhou Normal University, and the EU FP7 Grant 611272 (project GROWTHCOM).

\bibliography{Bibliography}

\end{document}